\documentclass[aps,nofootinbib,showpacs]{revtex4}
\usepackage{graphicx}

\begin{document}

\preprint{hep-th/0201012}

\title{Rotational Perturbations of Friedmann-Robertson-Walker Type
Brane-World Cosmological Models}

\author{Chiang-Mei Chen}
 \email{cmchen@phys.ntu.edu.tw}
\affiliation{Department of Physics, National Taiwan University,
Taipei 106, Taiwan}

\author{T. Harko}
 \email{tckarko@hkusua.hku.hk}
\affiliation{Department of Physics, The University of Hong Kong,
Pokfulam, Hong Kong}

\author{W. F. Kao}
 \email{wfgore@cc.nctu.edu.tw}
\affiliation{Institute of Physics, Chiao Tung University, Hsinchu,
Taiwan}

\author{M. K. Mak}
 \email{mkmak@vtc.edu.hk}
\affiliation{Department of Physics, The Hong Kong University of
Science and Technology, Clear Water Bay, Hong Kong}

\date{April 16, 2002}

%% REVTEX4
%\maketitle

\begin{abstract}
First order rotational perturbations of the
Friedmann-Robertson-Walker metric are considered in the framework
of the brane-world cosmological models. A rotation equation,
relating the perturbations of the metric tensor to the angular
velocity of the matter on the brane is derived under the
assumption of slow rotation. The mathematical structure of the
rotation equation imposes strong restrictions on the temporal and
spatial dependence of the brane matter angular velocity. The study
of the integrable cases of the rotation equation leads to three
distinct models, which are considered in detail. As a general
result we find that, similarly to the general relativistic case,
the rotational perturbations decay due to the expansion of the
matter on the brane. One of the obtained consistency conditions
leads to a particular, purely inflationary brane-world
cosmological model, with the cosmological fluid obeying a
non-linear barotropic equation of state.
\end{abstract}

%% REVTEX4

\maketitle

%%%%%%%%%%%%%%%%%%%%%%%%%%%%%%%%%%%%%%%%%%%%%%%%%%%%%%%%%%%%%%%%%%%%%%
\section{Introduction}
%%%%%%%%%%%%%%%%%%%%%%%%%%%%%%%%%%%%%%%%%%%%%%%%%%%%%%%%%%%%%%%%%%%%%%
On an astronomical scale rotation is a basic property of cosmic
objects. The rotation of planets, stars and galaxies inspired
Gamow to suggest that the Universe is rotating and the angular
momentum of stars and galaxies could be a result of the cosmic
vorticity \cite{Ga46}. But even that observational evidences of
cosmological rotation have been reported \cite{Bi82}-\cite{Ku97},
they are still subject of controversy.

From the analysis of microwave background anisotropy Collins and
Hawking \cite{CoHa73} and Barrow, Juszkiewicz and Sonoda
\cite{BaJuSo85} have found some very tight limits of the
cosmological vorticity, $T_{obs}>3\times 10^5 \, T_H$, where
$T_{obs}$ is the actual rotation period of our Universe and
$T_H=(1\sim 2)\times 10^{10}$ years is the Hubble time. Therefore
our present day Universe is rotating very slowly, if at all.

From a theoretical point of view in 1949 G\"odel \cite{Go49} gave
his famous example of a rotating cosmological solution to the
Einstein gravitational field equations. The G\"odel metric,
describing a dust Universe with energy density $\rho$ in the
presence of a negative cosmological constant $\Lambda$ is
\begin{equation}  \label{0}
ds^2 = \frac1{2\omega^2} \left[ -(dt + e^x dz)^2 + dx^2 + dy^2 +
\frac12 e^{2x} dz^2 \right].
\end{equation}
In this model the angular velocity of the cosmic rotation is given
by $\omega^2=4\pi\rho=-\Lambda$. G\"odel also discussed the
possibility of a cosmic explanation of the galactic rotation
\cite{Go49}. This rotating solution has attracted considerable
interest because the corresponding Universes possess the property
of closed time-like curves.

The investigation of rotating and rotating-expanding Universes
generated a large amount of literature in the field of general
relativity, the combination of rotation with expansion in
realistic cosmological models being one of the most difficult
tasks in cosmology (see \cite{Ob00} for a recent review of the
expansion-rotation problem in general relativity). Hence rotating
solutions of the gravitational field equations cannot be excluded
{\em a priori}. But this raises the question of why the Universe
rotates so slowly. This problem can also be naturally solved in
the framework of the inflationary model. Ellis and Olive
\cite{ElOl83} and Gr{\o}n and Soleng \cite{GrSo87} pointed out
that if the Universe came into being as a mini-universe of Planck
dimensions and went directly into an inflationary epoch driven by
a scalar field with a flat potential, due to the non-rotation of
the false vacuum and the exponential expansion during inflation
the cosmic vorticity has decayed by a factor of about $10^{-145}$.
The most important diluting effect of the order of $10^{-116}$ is
due to the relative density of the rotating fluid compared to the
non-rotating decay products of the false vacuum \cite{GrSo87}.
Inflationary cosmology also ruled out the possibility that the
vorticity of galaxies and stars be of cosmic origin.

The possibility of incorporating a slowly rotating Universe into
the framework of Friedmann-Robertson-Walker (FRW) type metrics has
been considered by Bayin and Cooperstock \cite{BaCo80}, who
obtained the restrictions imposed by the field equations on the
matter angular velocity. They also shown that uniform rotation is
incompatible with the dust filled (zero pressure) and with the
radiation dominated Universe. Bayin \cite{Ba85} has also shown
that the field equations admit solutions for a special class of
nonseparable rotation functions of the matter distribution. The
investigation of the first order rotational perturbations of flat
FRW type Universes proved to be useful in the study of string
cosmological models with dilaton and axion fields \cite{ChHaMa01}.
The form of the rotation equation imposes strong constraints on
the form of the dilaton field potential $U$, restricting the
allowed forms to two: the trivial case $U=0$ and the exponential
type potential.

Recently, Randall and Sundrum \cite{RS99a,RS99b} have pointed out
that a scenario with an infinite fifth dimension in the presence
of a brane can generate a theory of gravity which mimics purely
four-dimensional gravity, both with respect to the classical
gravitational potential and with respect to gravitational
radiation. The gravitational self-couplings are not significantly
modified in this model. This result has been obtained from the
study of a single 3-brane embedded in five dimensions, with the 5D
metric given by $ds^2=e^{-f(y)} \eta_{\mu\nu}dx^\mu dx^\nu+dy^2$,
which can produce a large hierarchy between the scale of particle
physics and gravity due to the appearance of the warp factor. Even
if the fifth dimension is uncompactified, standard 4D gravity is
reproduced on the brane. In contrast to the compactified case,
this follows because the near-brane geometry traps the massless
graviton. Hence this model allows the presence of large or even
infinite non-compact extra dimensions. Our brane is identified to
a domain wall in a 5-dimensional anti-de Sitter space-time.

The Randall-Sundrum (RS) model was inspired by superstring theory.
The ten-dimensional $E_8 \times E_8$ heterotic string theory,
which contains the standard model of elementary particle, could be
a promising candidate for the description of the real Universe.
This theory is connected with an eleven-dimensional theory
compactified on the orbifold $R^{10} \times S^1/Z_2 $ \cite{HW96}.
In this model we have two separated ten-dimensional manifolds.

The static RS solution has been extended to time-dependent
solutions and their cosmological properties have been extensively
studied \cite{KK00}-\cite{AnNuOl00} (for a review of dynamics and
geometry of brane universes see \cite{Ma01}).

The effective gravitational field equations on the brane world, in
which all the matter forces except gravity are confined on the
3-brane in a 5-dimensional space-time with $Z_2$-symmetry have
been obtained, by using an elegant geometric approach, by
Shiromizu, Maeda and Sasaki \cite{SMS00,SSM00}. The correct
signature for gravity is provided by the brane with positive
tension. If the bulk space-time is exactly anti-de Sitter,
generically the matter on the brane is required to be spatially
homogeneous. The electric part of the 5-dimensional Weyl tensor
$E_{IJ}$ gives the leading order corrections to the conventional
Einstein equations on the brane. The effect of the dilaton field
in the bulk can also be taken into account in this approach
\cite{MW00}.

The linearized perturbation equations in the generalized RS model
have been obtained, by using the covariant nonlinear dynamical
equations for the gravitational and matter fields on the brane, by
Maartens \cite{Ma00}. The behavior of an anisotropic Bianchi type
I brane-world in the presence of inflationary scalar fields has
been considered by Maartens, Sahni and Saini \cite{MSS00}. A
systematic analysis, using dynamical systems techniques, of the
qualitative behavior of the FRW, Bianchi type I and V cosmological
models in the RS brane world scenario, with matter on the brane
obeying a barotropic equation of state has been performed by
Campos and Sopuerta \cite{CS01,CS01a}. In particular, they
constructed the state spaces for these models and discussed what
new critical points appear, the occurrence of bifurcations and the
dynamics of the anisotropy. The general exact solution of the
field equations for an anisotropic brane with Bianchi type I and V
geometry, with perfect fluid and scalar fields as matter sources
has been found in \cite{ChHaMa01a}. Expanding Bianchi type I and V
brane-worlds always isotropize, although there could be
intermediate stages in which the anisotropy grows. In spatially
homogeneous brane world cosmological models the initial
singularity is isotropic and hence the initial conditions problem
is solved \cite{Co01a}. Consequently, these models do not exhibit
Mixmaster or chaotic-like behavior close to the initial
singularity \cite{Co01b}.

Realistic brane-world cosmological models require the
consideration of more general matter sources to describe the
evolution and dynamics of the very early Universe. Hence the
effect of the bulk viscosity of the matter on the brane have been
analyzed in \cite{ChHaMa01b}. Limits on the initial anisotropy
induced by the 5-dimensional Kaluza-Klein graviton stresses by
using the CMB anisotropies have been obtained by Barrow and
Maartens \cite{BaMa01}. Anisotropic Bianchi type I brane-worlds
with a pure magnetic field and a perfect fluid have also been
analyzed \cite{BaHe01}.

It is the purpose of the present paper to investigate the effects
of the rotational perturbations on a brane world with FRW type
geometry. A similar case was also considered in \cite{BMW01}.
Assuming the rotation is slow and by keeping only the first order
rotational terms in the field equations, a rotation equation
describing the time and space evolution of the metric
perturbations is obtained. This equation also contains the
angular velocity of the matter rotating on the brane. By assuming
that the metric perturbation and the matter angular velocity are
separable functions of the variables $t$ and $r$, the
mathematical consistency of the rotation equation leads to some
restrictions on the functional form of the angular velocity. In
particular a class of solutions of the field equations leads to a
barotropic brane-world cosmological model, with a non-linear
pressure-energy density dependence. But generally, similar to the
general relativistic case, the rotational perturbations will
rapidly decay due to the expansion of the Universe. However, this
general result is valid only in the presence of the dark energy
term, describing the influence of the five-dimensional bulk on
the brane. If this term is zero, rotational perturbations in a
very high density (stiff) cosmological fluid decay only in the
so-called case of the ``perfect dragging''.

The present paper is organized as follows. The field equations for
a slowly rotating brane-world are written down and the basic
rotation equation is obtained in Section II. The integrable cases
of the rotation equation are considered in Section III. A brane
world cosmological model, derived from the mathematical
consistency requirement of the rotation equation, is obtained in
Section IV. In Section V we discuss and conclude our results.

%%%%%%%%%%%%%%%%%%%%%%%%%%%%%%%%%%%%%%%%%%%%%%%%%%%%%%%%%%%%%%%%%%%%%%
\section{Geometry, Brane-World Field Equations and Consequences}
%%%%%%%%%%%%%%%%%%%%%%%%%%%%%%%%%%%%%%%%%%%%%%%%%%%%%%%%%%%%%%%%%%%%%%
In the 5D space-time the brane-world is located as $Y(X^I)=0$,
where $X^I, \, I=0,1,2,3,4$ are 5-dimensional coordinates. The
effective action in five dimensions is \cite{MW00}
\begin{equation}
S = \int d^5X \sqrt{-g_5} \left( \frac1{2k_5^2} R_5 - \Lambda_5
\right) + \int_{Y=0} d^4x \sqrt{-g} \left( \frac1{k_5^2} K^\pm -
\lambda + L^{\text{matter}} \right),
\end{equation}
with $k_5^2=8\pi G_5$ the 5-dimensional gravitational coupling
constant and where $x^\mu, \, \mu=0,1,2,3$ are the induced
4-dimensional brane world coordinates. We use a system of units so
that the speed of light $c=1$. $R_5$ is the 5D intrinsic curvature
in the bulk and $K^\pm$ is the extrinsic curvature on either side
of the brane.

On the 5-dimensional space-time (the bulk), with the negative
vacuum energy $\Lambda_5$ as only source of the gravitational
field the Einstein field equations are given by
\begin{equation}
G_{IJ} = k_5^2 T_{IJ}, \qquad T_{IJ} = - \Lambda_5 g_{IJ} +
\delta(Y) \left[ -\lambda g_{IJ} + T_{IJ}^{\text{matter}} \right],
\end{equation}
In this space-time a brane is a fixed point of the $Z_2$ symmetry.
In the following capital Latin indices run in the range $0,...,4$
while Greek indices take the values $0,...,3$.

Assuming a metric of the form $ds^2=(n_I n_J + g_{IJ})dx^I dx^J$,
with $n_I dx^I = d\chi$ the unit normal to the
$\chi=\text{constant}$ hypersurfaces and $g_{IJ}$ the induced
metric on $\chi=\text{constant}$ hypersurfaces, the effective
four-dimensional gravitational equations on the brane take the
form \cite{SMS00,SSM00}:
\begin{equation}
G_{\mu\nu} = - \Lambda g_{\mu\nu} + k_4^2 T_{\mu\nu} + k_5^4
S_{\mu \nu} - E_{\mu\nu},
\end{equation}
where
\begin{equation}
S_{\mu\nu} = \frac1{12} T T_{\mu\nu} - \frac14 T_\mu{}^\alpha
T_{\nu\alpha} + \frac1{24} g_{\mu\nu} \left( 3 T^{\alpha\beta}
T_{\alpha\beta} - T^2 \right),
\end{equation}
and $\Lambda =k_5^2(\Lambda_5+k_5^2 \lambda^2/6)/2, \,
k_4^2=k_5^4\lambda/6$ and $E_{IJ}=C_{IAJB}n^A n^B$. $C_{IAJB}$ is
the 5-dimensional Weyl tensor in the bulk and $\lambda$ is the
vacuum energy on the brane. $T_{\mu\nu}$ is the matter
energy-momentum tensor on the brane and $T=T^\mu{}_\mu$ is the
trace of the energy-momentum tensor.

For any matter fields (scalar field, perfect fluids, kinetic
gases, dissipative fluids etc.) the general form of the brane
energy-momentum tensor can be covariantly given as
\begin{equation}
T_{\mu\nu} = (\rho+p) u_\mu u_\nu + p h_{\mu\nu} + \pi_{\mu\nu} +
2 q_{(\mu} u_{\nu)}.  \label{EMT}
\end{equation}
The decomposition is irreducible for any chosen 4-velocity
$u^\mu$. Here $\rho$ and $p$ are the energy density and isotropic
pressure, and $h_{\mu\nu}=g_{\mu\nu}+u_\mu u_\nu$ projects
orthogonal to $u^\mu$. The energy flux obeys $q_\mu=q_{<\mu>}$,
and the anisotropic stress obeys $\pi_{\mu\nu}=\pi_{<\mu\nu>}$,
where angular brackets denote the projected, symmetric and
tracefree part:
\begin{equation}
V_{<\mu>} = h_\mu{}^\nu V_\nu, \qquad W_{<\mu\nu>} = \left[
h_{(\mu}{}^\alpha h_{\nu)}{}^\beta - \frac13 h^{\alpha\beta}
h_{\mu\nu} \right] W_{\alpha\beta}.
\end{equation}

The symmetric properties of $E_{\mu\nu}$ imply that in general we
can decompose it irreducibly with respect to a chosen 4-velocity
field $u^\mu$ as
\begin{equation}
E_{\mu\nu} = -\frac6{\lambda k_4^2} \left[ {\cal U} \left( u_\mu
u_\nu + \frac13 h_{\mu\nu} \right) + {\cal P}_{\mu\nu} + 2 {\cal
Q}_{(\mu} u_{\nu)} \right],  \label{WT}
\end{equation}
where ${\cal U}$ is a scalar, ${\cal Q}_\mu$ a spatial vector and
${\cal P}_{\mu\nu}$ a spatial, symmetric and trace-free tensor.
For a FRW model ${\cal Q}_\mu={\cal P}_{\mu\nu}=0$ \cite{CS01a}
and hence the only non-zero contribution from the 5-dimensional
Weyl tensor from the bulk is given by the scalar term ${\cal U}$.

The Einstein equation in the bulk imply the conservation of the
energy momentum tensor of the matter on the brane,
\begin{equation}
T_\mu{}^\nu{}_{;\nu} \mid_{\chi=0} = 0.
\end{equation}

The rotationally perturbed metric can be expressed in terms of the
usual coordinates in the form \cite{MuFeBr92}
\begin{equation}
ds^2 = - dt^2 + a^2(t) \left[ \frac{dr^2}{1-kr^2} + r^2 \left(
d\theta^2 + \sin^2\theta d\varphi^2 \right) \right] - 2\Omega(t,r)
a^2(t) r^2 \sin^2\theta \,dt d\varphi, \label{R6}
\end{equation}
where $\Omega(t,r)$ is the metric rotation function. Although
$\Omega$ plays a role in the ``dragging'' of local inertial
frames, it is not the angular velocity of these frames, except for
the special case when it coincides with the angular velocity of
the matter fields. $k=1$ corresponds to closed Universes, with
$0\leq r\leq 1$. $k=-1$ corresponds to open Universes, while the
case $k=0$ describes a flat geometry, where the range of $r$ is
$0\leq r<\infty$. In all models, the time-like variable $t$ ranges
from $0$ to $\infty$.

For the matter energy-momentum tensor on the brane we restrict our
analysis to the case of the perfect fluid energy-momentum tensor,
\begin{equation}
T^{\mu\nu} = (\rho+p) u^\mu u^\nu + p g^{\mu\nu}.
\end{equation}
The components of the four-velocity vector are $u^0=1, \,
u^1=u^2=0$ and $u^3=\omega(t,r)$. $\omega=d\varphi/dt$ is the
angular velocity of the matter distribution. Consequently, for the
rotating brane the energy-momentum tensor has a supplementary
component
\begin{equation}
T_{03} = \left\{ \left[ \Omega(t,r)-\omega(t,r) \right] \rho -
\omega(t,r) p \right\} r^2 a^2(t) \sin^2\theta.
\end{equation}

We assume that rotation is sufficiently slow so that deviations
from spherical symmetry can be neglected. Then to first order in
$\Omega$ the gravitational and field equations become
\begin{eqnarray}
3\frac{\dot a^2}{a^2} + \frac{3k}{a^2} &=& \Lambda + k_4^2 \rho +
\frac{k_4^2}{2\lambda} \rho^2 + \frac6{\lambda k_4^2}{\cal U},
\label{dH} \\
\frac{2\ddot a}{a} + \frac{\dot a^2}{a^2} + \frac{k}{a^2} &=&
\Lambda - k_4^2 p - \frac{k_4^2}{\lambda} \rho p -
\frac{k_4^2}{2\lambda} \rho^2 - \frac2{\lambda k_4^2}{\cal U},
\label{dVHi} \\
\dot \rho + 3 (\rho+p) \frac{\dot a}{a} &=& 0,  \label{drho} \\
\dot {\cal U} + \frac{4 \dot a}{a} {\cal U} &=& 0, \label{u}
\end{eqnarray}
\begin{eqnarray}
3\frac{\dot a}{a} \frac{\partial \Omega(t,r)}{\partial r} +
\frac{\partial^2 \Omega(t,r)}{\partial t\partial r} &=& 0,
\label{r13} \\
(1-kr^2) \frac{\partial^2 \Omega(t,r)}{\partial r^2} + \left(
\frac4{r} - 5kr \right) \frac{\partial \Omega(t,r)}{\partial r} -
4 \left[ \Omega(t,r) - \omega(t,r) \right] (k - a \ddot a + \dot
a^2) &=& 0.  \label{r03}
\end{eqnarray}
The last two equations follows from the $R_{13}$ and $R_{03}$
components of the field equations, respectively. Generally, we
shall assume that the thermodynamic pressure $p$ and the energy
density $\rho$ are related by a barotropic equation of state
$p=p(\rho)$.

From a mathematical point of view the field equations
(\ref{dH})-(\ref{dVHi}) and (\ref{u})-(\ref{r03}) represent a
system of five equations in five unknowns $a(t), \rho(t), {\cal
U}, \Omega(t,r)$ and $\omega(t,r)$ (the Bianchi identity
(\ref{drho}) is a consequence of the field equations and the
pressure can be eliminated via the equation of state of the matter
on the brane). Therefore to find the general solution of this
system is a well-posed problem. Since the Bianchi identity and the
evolution equation for ${\cal U}$ can be generally integrated,
giving the energy density $\rho$ and the dark matter term as some
functions of the scale factor $a$, the basic field equations
describing the first order rotational perturbations of brane world
cosmologies are Eqs. (\ref{dH}) and (\ref{r13})-(\ref{r03}),
three equations in three unknowns $a(t), \Omega(t,r)$ and
$\omega(t,r)$.

Mathematically, Eqs. (\ref{r13})-(\ref{r03}) represent a system
of partial differential equations. To obtain the solution of these
equations we shall use the standard technique of the separation of
variables \cite{Bo83}. Therefore, we generally assume that both
the rotation metric function and the angular velocity of the
matter on the brane are the product of two functions, the first
one depending on the cosmological time $t$ only and the second one
on the radial variable $r$ only. When these representations of the
unknown functions are substituted back to Eqs.
(\ref{r13})-(\ref{r03}), the field equations become identities in
the independent variable, which can be separated in two
independent equations, giving the time evolution and the spatial
behavior of the considered physical parameters. If the values of
the separation constants are properly chosen, then the combination
of the solutions of the separated equations gives the solution of
the initial partial differential equation for any boundary or
initial conditions \cite{Bo83}.

Eq. (\ref{u}) can be immediately integrated to give the following
general expression for the ``dark energy'' ${\cal U}$:
\begin{equation}
{\cal U} = \frac{{\cal U}_0}{a^4},
\end{equation}
with ${\cal U}_0>0$ a constant of integration.

The temporal and spatial dependence of $\Omega(t,r)$ is determined
by Eqs. (\ref{r13}) and (\ref{r03}). The general solution of Eq.
(\ref{r13}) can be immediately found and is given by
\begin{equation}
\Omega(t,r) = \frac{A(r)}{a^3(t)},  \label{omega}
\end{equation}
where $A(r)$ is a function to be determined from the field
equations and an arbitrary time dependent function has been set to
zero, without altering the physical structure of the model.
Substituting Eq. (\ref{omega}) into Eq. (\ref {r03}) gives the
following equation describing the evolution of the rotational
perturbations on the brane:
\begin{equation}
(1-kr^2) \frac{d^2 A(r)}{dr^2} + \left( \frac4{r} - 5kr \right)
\frac{dA(r)}{dr} - 4 \left[ A(r) - \omega(t,r) a^3 \right]
(k-a\ddot{a}+\dot{a}^2) = 0. \label{eq}
\end{equation}

By introducing a new variable $\eta=r^2$, Eq. (\ref{eq}) can be
transformed into the following form:
\begin{equation}
\eta ( 1 - k \eta ) \frac{d^2 A(\eta)}{d\eta^2} + \left( \frac52 -
3 k \eta \right) \frac{dA(\eta)}{d\eta} = \left[ A(\eta) -
\omega(t,\eta) a^3 \right] (k-a\ddot{a}+\dot{a}^2). \label{eqfin}
\end{equation}

Eq. (\ref{eqfin}), the basic rotation equation, governs the
evolution of the first order rotational perturbations of brane
world models. For a given equation of state of the matter, $a(t)$
is obtained from the field equations (\ref{dH})-(\ref{u}). Hence
Eq. (\ref{eqfin}) places some restrictions on the possible form of
the angular velocity of the matter $\omega(t,r)$. Indeed, since
the left hand side of the Eq. (\ref{eqfin}) is a function of $r$
alone, the right hand side must be either a function of $r$ alone
or a function of $t$ alone (and hence set equal to a constant). As
a first consequence of the mathematical structure of Eq.
(\ref{eqfin}) it follows that the angular velocity of the rotating
brane world must also be a separable function and we assume it to
be of the form $\omega(t,r)=f(t)g(r)$, with $f(t)$ and $g(r)$
functions to be determined from Eq. (\ref{eqfin}).

%%%%%%%%%%%%%%%%%%%%%%%%%%%%%%%%%%%%%%%%%%%%%%%%%%%%%%%%%%%%%%%%%%%%%%
\section{Consistency conditions for the rotation equation}
%%%%%%%%%%%%%%%%%%%%%%%%%%%%%%%%%%%%%%%%%%%%%%%%%%%%%%%%%%%%%%%%%%%%%%
The first and simplest case in which Eq. (\ref{eqfin}) has a
solution corresponds to the case $\omega(t,r)=\Omega(t,r)$.
Physically, this situation corresponds to the so-called ``perfect
dragging'', in which the function $\Omega(t,r)$ appearing in the
rotationally perturbed line element Eq. (\ref{R6}) is the angular
velocity of the rotating matter. Then the spatial evolution of the
angular velocity is determined from the equation
\begin{equation}
\eta (1 - k \eta) \frac{d^2 A(\eta)}{d\eta^2} + \left( \frac52 - 3
k \eta \right) \frac{dA(\eta)}{d\eta} = 0. \label{case1}
\end{equation}

Eq. (\ref{case1}) has the following solutions:
%\begin{eqnarray}
%A(r) &=& C_1^{(1)} (1 - r^2)^{1/2} \left( \frac2{r} + \frac1{r^3}
%\right) + C_2^{(1)}, \qquad k=+1, \\
%A(r) &=& C_1^{(-1)} (1 + r^2)^{1/2} \left( \frac2{r} - \frac1{r^3}
%\right) + C_2^{(-1)}, \qquad k=-1, \\
%A(r) &=& \frac{C_1^{(0)}}{r^3} + C_2^{(0)}, \qquad\qquad k=0,
%\end{eqnarray}
\begin{equation}
A(r) = \left\{ \begin{array}{ll}
 C_1^{(1)} (1 - r^2)^{1/2} \left( 2r^{-1} + r^{-3}
\right) + C_2^{(1)}, & \qquad k=+1, \\
 C_1^{(-1)} (1 + r^2)^{1/2} \left( 2r^{-1} - r^{-3}
\right) + C_2^{(-1)}, & \qquad k=-1, \\
 C_1^{(0)} r^{-3} + C_2^{(0)}, &\qquad k=0,
 \end{array} \right.
\end{equation}
where $C_1^{(i)}$ and $C_2^{(i)}, \, i=-1,0,1$ are arbitrary
constants of integration.

These solutions are not regular at the origin. They could be used
in regions away from the origin and joined continuously to other
solutions which are regular at the origin, which could be obtained
by taking into account second order or higher terms in $\Omega$ in
the field equations. The temporal behavior of the angular velocity
is given by $a^{-3}(t)$. Thus for an expanding Universe this type
of dependence guarantees the decay of the rotational perturbations
in the limit of large cosmological times.

The second case in which the rotation equation can be solved
corresponds to the choice $g(r)=A(r)$. In this case both
$\omega(t,r)$ and $\Omega(t,r)$ have the same spatial dependence,
but their time dependence is different. Therefore Eq.
(\ref{eqfin}) can be decoupled into the following two simple
ordinary differential equations:
\begin{eqnarray}
\eta (1 - k \eta) \frac{d^2 A(\eta)}{d\eta^2} + \left( \frac52 - 3
k \eta \right) \frac{d A(\eta)}{d\eta} - C A(\eta) &=& 0,
\label{case2r} \\
\left[ 1 - f(t) a^3 \right] (k-a\ddot{a}+\dot{a}^2) &=& C,
\label{case2t}
\end{eqnarray}
where $C$ is a separation constant.

Consider first the time dependence of the angular velocity. With
the use of the field equations Eqs. (\ref{dH})-(\ref{u}), Eq.
(\ref{case2t}) gives
\begin{equation}
f(a) = \frac{a^2 \left[ k_4^2 (\rho+p) + (k_4^2/\lambda) \rho^2 +
(k_4^2/\lambda) \rho p + (8{\cal U}_0/\lambda k_4^2) a^{-4}
\right] - 2 C}{a^5 \left[ k_4^2 (\rho+p) + (k_4^2/\lambda) \rho^2
+ (k_4^2/\lambda) \rho p + (8{\cal U}_0/\lambda k_4^2) a^{-4}
\right]}. \label{fcase2}
\end{equation}

In the very early stages of the evolution of the Universe it is
natural to assume that the very high density matter obeys a
barotropic equation of state of the form $p=(\gamma-1)\rho$, with
$\gamma$ a constant and $1\leq \gamma \leq 2$. $\gamma=2$
corresponds to the extreme limit of very high densities in which
the speed of sound equals the speed of light (stiff case)
\cite{ShTe83}. For a barotropic equation of state the conservation
equation Eq. (\ref{drho}) can immediately be integrated to give
$\rho=\rho_0 a^{-3\gamma}$.

Therefore for a cosmological fluid obeying a barotropic equation
of state the time dependence of the angular velocity of the slowly
rotating brane world is given by
\begin{equation}
f(a) = \frac{\gamma \rho_0 k_4^2 a^2 \left[ a^{3\gamma} + (8{\cal
U}_0/\gamma \lambda k_4^4 \rho_0) a^{6\gamma-4} + (\rho_0/\lambda)
\right] - 2 C a^{6\gamma}}{\gamma \rho_0 k_4^2 a^5 \left[
a^{3\gamma} + (8{\cal U}_0/\gamma \lambda k_4^4 \rho_0)
a^{6\gamma-4} + (\rho_0/\lambda) \right]}.
\end{equation}

The variation of the function $f(a)$ for different values of
$\gamma$ is represented in Fig. 1.

%% REVTEX4
%%%%%%%%%%%%%%%%%%%%%%%%%%%%%%%%%%%%%%%%%%%%%%%%%%%%%%%%%%%%%%%
\begin{figure}
\includegraphics[width=10cm]{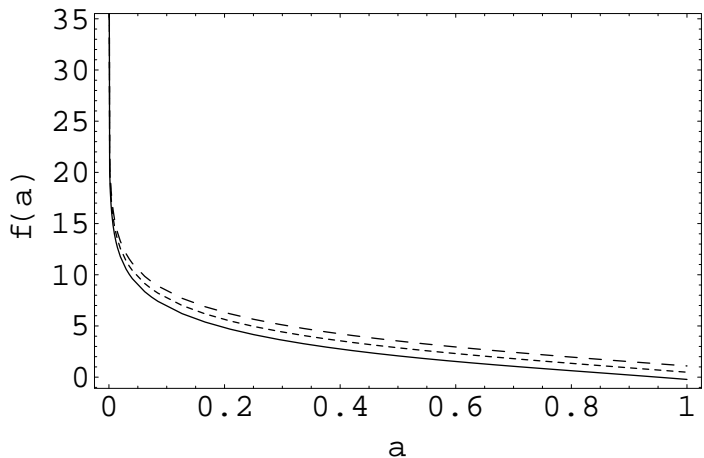}
\caption{Dynamics of the temporal part of the angular velocity
$f(a)$ (in a logarithmic scale) as a function of the scale factor
$a$ for different equations of state of the brane matter:
$\gamma=2$ (solid curve), $\gamma=4/3$ (dotted curve) and
$\gamma=1$ (dashed curve). We have normalized the parameters so
that $\rho_0=\lambda, \, 8 {\cal U}_0=k_4^4 \rho_0^2$ and
$2C=k_4^2 \rho_0$.}
\end{figure}
%%%%%%%%%%%%%%%%%%%%%%%%%%%%%%%%%%%%%%%%%%%%%%%%%%%%%%%%%%%%%%%

For all values of $\gamma$ and for an expanding Universe the
rotational perturbations tend, in the large time limit, to zero.
This result is also independent on the numerical value of the
separation constant $C$.

In the particular case of the vanishing dark matter, ${\cal
U}_0=0$, and in the limit of large times $f(a)$ tends to the limit
\begin{equation}
f(a) = a^{-3} - \frac{2C}{\gamma \rho _0 k_4^2} a^{3\gamma-5}.
\end{equation}

Therefore in the absence of dark energy the rotational
perturbations on the brane would decay in time only if the
barotropic fluid satisfies the condition $\gamma<5/3$. In
particular this condition is not satisfied by the stiff
cosmological fluid with $\gamma=2$. Hence in this case the
observational evidence of a non-rotating Universe requires the
condition $C=0$ and, consequently, a ``perfect dragging'' of the
cosmological fluid. But in a radiation fluid with $\gamma=4/3$ the
rotational perturbations will decay in time.

For this model the spatial dependence of the angular velocity is
described by Eq. (\ref{case2r}). For $k=+1$ this equation is of
the form $x(1-x) F_{xx} + \left[\delta-(1+\alpha+\beta)x\right]
F_{x} - \alpha\beta F=0$, (the hypergeometric equation), with
$\delta \neq 0,1,2...$ and $\alpha, \beta$ are constants. The
general solution of the hypergeometric equation is given by $F=C_1
F(\alpha,\beta;\delta;x)+C_2 x^{1-\delta } F(1-\delta+\alpha,
1-\delta+\beta;2-\delta;x)$, where $C_1$ and $C_2$ are arbitrary
constants of integration, and $F(\alpha,\beta;\delta;x) =
\sum_{n=0}^{\infty} \frac{(\alpha)_n (\beta)_n}{n! (\delta)_n}
x^n$, where $(\alpha)_0=1, \, (\alpha)_{n} = \Gamma(\alpha+n) /
\Gamma(\alpha) = \alpha(\alpha +1)...(\alpha+n-1), n=1,2,...$
\cite{AbSt72}. The radius of convergence of the solution is unity
and if one of the constants $\alpha, \beta, \delta-\alpha,
\delta-\beta$ is a negative integer, then the series terminates
\cite{AbSt72}.

For Eq. (\ref{case2r}) $\delta=5/2$ and the constants $\alpha$ and
$\beta$ must satisfy the conditions $\alpha+\beta=2$ and $\alpha
\beta=C$. The separation constant $C$ must satisfy the condition
$C\le 1$.

Therefore the general solution of Eq. (\ref{case2r}) is given by
\begin{equation}
g(r) = A(r) = C_1^{(1)} \sum_{n=0}^{\infty} \frac{(\alpha)_n
(\beta)_n}{n! \left( \frac52 \right)_n} \, r^{2n} + C_2^{(1)}
r^{-3} \sum_{n=0}^{\infty} \frac{ \left( \alpha-\frac32 \right)_n
\left( \beta-\frac32 \right)_n}{n! \left( -\frac12 \right)_{n}} \,
r^{2n}, \qquad k=+1.
\end{equation}
Since the second term is not regular at $r=0$ we must take
$C_2^{(1)}=0$. Therefore the resulting solution is also defined
and is regular at the origin of the radial coordinate.

For open models $k=-1$ and the solution of Eq. (\ref{case2r}) can
be obtained from the previous one by replacing $\eta$ by $-\eta$
and the separation constant $C$ by $-C$. These solutions are
convergent only for $\eta<1$. For solutions regular outside
$\eta=1$, one must consider solutions found in the neighborhood of
the remaining two singular points of the differential equation,
namely $1$ and $\infty$ \cite{AbSt72}.

For $k=0$ Eq. (\ref{eq}) gives
\begin{equation}
\frac{d^2 A(r)}{dr^2} + \frac4{r} \frac{dA(r)}{dr} - 4CA(r) = 0,
\end{equation}
with the general solution given by \cite{ChHaMa01}
\begin{equation}
g(r) = A(r) = \frac1{r^2} \left[ 2\sqrt{C} \left( C_1^{(0)}
e^{2\sqrt{C} r} + C_2^{(0)} e^{-2\sqrt{C} r} \right) - \frac1{r}
\left( C_1^{(0)} e^{2\sqrt{C} r} - C_2^{(0)} e^{-2\sqrt{C} r}
\right) \right], \qquad k = 0,
\end{equation}
and $C_1^{(0)}$ and $C_2^{(0)}$ arbitrary constants. Hence we have
obtained in a closed form all the possible spatial distribution
functions corresponding to the time dependence Eq. (\ref{fcase2})
of the angular velocity of the slowly rotating brane world.

The third case in which the variable in Eq. (\ref{eq}) can be
separated is given by assuming that $\omega(t,r)=G(r)/a^3(t)$ and
the time only dependent term in the right hand side of the
equation is a constant. Therefore we obtain:
\begin{eqnarray}
\eta (1 - k \eta) \frac{d^2 A(\eta)}{d\eta^2} + \left( \frac52 - 3
k \eta \right) \frac{d A(\eta)}{d\eta} - K \left[ A(\eta) -
G(\eta) \right] &=& 0,  \label{case3r} \\
k - a\ddot{a} + \dot{a}^{2} &=& K,  \label{case3t}
\end{eqnarray}
where $K$ is a separation constant. Generally Eq. (\ref{case3r})
cannot be solved, due to a lack of exact knowledge of the
mathematical form of the function $G(\eta)$. In the particular
case $G(\eta)=G_0=\text{constant}$, the solutions of Eq.
(\ref{case3r}) can be obtained by the substitution $A(r) \to A(r)
+ G_0$, where $A(r)$ is any of the solutions of the homogeneous
equation Eq. (\ref{case2r}).

Finally, we consider the possibility of the existence of an
uniformly rotating brane world, with
$\omega=\omega_{0}=\text{constant}$. In this case the rotation
equation takes the form
\begin{equation}
\eta (1 - k \eta) \frac{d^2 A(\eta)}{d\eta^2} + \left( \frac52 - 3
k \eta \right) \frac{d A(\eta)}{d\eta} -
A(\eta)(k-a\ddot{a}+\dot{a}^{2}) = - \omega_0
a^3(k-a\ddot{a}+\dot{a}^{2}). \label{constrot}
\end{equation}

Therefore the mathematical consistency of Eq. (\ref{constrot})
requires either $A(\eta)=\text{constant}$, implying
$a=\text{constant}$ or $(k-a\ddot{a}+\dot{a}^2)=\text{constant}$
and $a^3(k-a\ddot{a}+\dot{a}^2)=\text{constant}$, also leading to
$a=\text{constant}$. Hence uniform rotation is possible only for a
static brane, the expansion of the Universe making the angular
velocity a complicated function of the temporal and spatial
coordinates.

%%%%%%%%%%%%%%%%%%%%%%%%%%%%%%%%%%%%%%%%%%%%%%%%%%%%%%%%%%%%%%%%%%%%%%
\section{An inflationary brane world cosmological model}
%%%%%%%%%%%%%%%%%%%%%%%%%%%%%%%%%%%%%%%%%%%%%%%%%%%%%%%%%%%%%%%%%%%%%%
In the previous Section we have shown that the third condition of
integrability of the rotation equation leads to the Eq.
(\ref{case3t}), describing the evolution of the scale factor of
the slowly rotating brane world. Hence in this case the time
evolution of the brane Universe is strongly correlated with its
rotational properties.

With the help of the substitutions $\dot a=u, \, u^2=v$, Eq.
(\ref{case3t}) can be transformed into a first order linear
differential equation of the form
\begin{equation}
\frac{dv}{da} = \frac2{a} v - \frac2{a} B,  \label{eqv}
\end{equation}
where we denoted $B=K-k$.
The general solution of Eq. (\ref{eqv})
is $v=H_0^2 a^2+B$, with $H_0^2$ a positive-definite arbitrary
constant of integration. We will assume $H_0 >0$ for later
convenience.

Therefore the general solution of Eq. (\ref{case3t}) is given by:
%\begin{eqnarray}
%a &=& \frac{\sqrt{B}}{H_0} \sinh\left[ H_0(t-t_0) \right], \qquad
%B>0, \label{a1} \\
%a &=& \frac{\sqrt{|B|}}{H_0} \cosh\left[ H_0(t-t_0) \right],
%\qquad B<0, \label{a2} \\
%a &=& e^{H_0(t-t_0)}, \qquad\qquad B=0, \label{a3}
%\end{eqnarray}
\begin{equation}\label{aa}
a = \left\{ \begin{array}{ll}
 \sqrt{B} H_0^{-1} \sinh\left[ H_0(t-t_0) \right], & \qquad
B>0, \\
 \sqrt{|B|} H_0^{-1} \cosh\left[ H_0(t-t_0) \right], & \qquad
B<0, \\
 e^{H_0(t-t_0)}, & \qquad B=0,
 \end{array} \right.
\end{equation}
where $t_0$ is a constant of integration.

In this class of models the Hubble parameter is given by
%\begin{eqnarray}
%H &=& H_0 \coth\left[ H_0(t-t_0) \right], \qquad B>0, \\
%H &=& H_0 \tanh\left[ H_0(t-t_0) \right], \qquad B<0, \\
%H &=& H_0 = \text{constant}, \qquad\qquad B=0.
%\end{eqnarray}
\begin{equation}
H = \left\{ \begin{array}{ll}
 H_0 \coth\left[ H_0(t-t_0) \right], &\qquad B>0, \\
 H_0 \tanh\left[ H_0(t-t_0) \right], &\qquad B<0, \\
 H_0 = \text{constant}, &\qquad B=0.
 \end{array} \right.
\end{equation}
The energy density and pressure of the matter on the brane follow
from the field equations Eqs. (\ref{dH}) and (\ref{dVHi}), with
the energy-density of the cosmological fluid given by
\begin{equation}
\rho = \lambda \left( \sqrt{\frac{6K}{\lambda k_4^2}\frac1{a^2} -
\frac{12{\cal U}_0}{\lambda^2 k_4^4}\frac1{a^4} + 1 -
\frac{2\Lambda}{\lambda k_4^2} + \frac{6H_0^2}{\lambda k_4^2}} - 1
\right). \label{rho3}
\end{equation}
In order to obtain Eq. (\ref{rho3}) we have used the identity
$3\dot a^2/a^2 + 3k/a^2 = 3K/a^2 + 3H_0^2$.

Due to the presence of dark matter term, the energy density of the
brane world has a maximum, $d\rho/da=0$, corresponding to a value
of the scale factor given by $a_{\max}=(2/k_4)\sqrt{{\cal
U}_0/K\lambda}$. For this value of $a$ the energy density
satisfies the condition $d^2 \rho/da^2<0$. The maximum value of
the energy density is given by $\rho_{\max}=\lambda \left(
\sqrt{3K^2/4{\cal U}_0 + 1 - 2\Lambda/\lambda k_4^2 +
6H_0^2/\lambda k_4^2} - 1 \right)$. If ${\cal U} = 0$ and the
scale factor $a$ is a monotonically increasing function of time,
then the energy density of the brane world is a monotonically
decreasing function for all times.

The condition of the non-negativity of the energy density $\rho
\geq 0$ can be reformulated as a condition on the scale factor of
the form $(\Lambda-3H_0^2) \lambda k_4^2 a^4 - 3 \lambda K k_4^2
a^2 + 6{\cal U}_0 < 0$, inequality that is satisfied for all $a$
if $3H_0^2>\Lambda$ and $3\lambda K^2 k_4^2 + 8(3H_0^2-\Lambda)
{\cal U}_0<0$. Since these two conditions cannot be satisfied
simultaneously, it follows that generally the energy density is
non-negative only for a finite time interval. For ${\cal U}_0=0$,
$\rho \geq 0$ for $a\in \left( 0,\sqrt{3K/(\Lambda-3H_0^2)}
\right]$.

The general pressure-energy density relation (the equation of
state) is given by
\begin{equation}\label{eqstate}
p = \frac{(2K/k_4^2)a^{-2} - (8{\cal U}_0/\lambda k_4^4) a^{-4} -
\rho - (\rho^2/\lambda)}{1+\rho/\lambda}.
\end{equation}
The condition of the non-negativity of the pressure requires the
condition
\begin{equation}
\frac{2K}{k_4^2} \frac1{a^2} - \frac{8{\cal U}_0}{\lambda k_4^4}
\frac1{a^4} \geq \rho \left( 1+\frac{\rho}{\lambda} \right),
\end{equation}
be satisfied for all $a$. From Eq. (\ref{dH}) we obtain
\begin{equation}
\rho \left( 1+\frac{\rho}{\lambda} \right) = \frac{6K}{k_4^2}
\frac1{a^2} - \frac{12{\cal U}_0}{\lambda k_4^4} \frac1{a^4} -
\rho + \frac{2(3H_0^2-\Lambda)}{k_4^2},
\end{equation}
and therefore the condition of the non-negativity of the pressure
can be reformulated as the following condition which must be
satisfied by the energy density of the matter on the brane:
\begin{equation}
\rho \geq \frac{4K}{k_4^2} \frac1{a^2} - \frac{4{\cal U}_0}{\lambda
k_4^4} \frac1{a^4} + \frac{2(3H_0^2-\Lambda)}{k_4^2}.
\end{equation}

Hence with the use of Eqs. (\ref{aa}) we obtain the following
exact analytic representations for the energy density and
pressure:
\begin{eqnarray}
\rho(t) &=& \lambda \left\{ \sqrt{a_0 \sinh^{-2}\left[ H_0(t-t_0)
\right] - b_0 \sinh^{-4}\left[ H_0(t-t_0) \right] + c_0} - 1
\right\}, \qquad B>0, \label{solut1} \\
p(t) &=& \frac{k_0 \sinh^{-2}\left[ H_0(t-t_0) \right] - u_0
\sinh^{-4}\left[ H_0(t-t_0) \right]}{\sqrt{a_0 \sinh^{-2}\left[
H_0(t-t_0) \right] - b_0 \sinh^{-4}\left[ H_0(t-t_0) \right] +
c_0}}  \nonumber \\
&-&\lambda \left\{ \sqrt{a_0 \sinh^{-2}\left[ H_0(t-t_0) \right] -
b_0 \sinh^{-4} \left[ H_0(t-t_0) \right] + c_0} - 1 \right\},
\qquad B>0,
\end{eqnarray}
\begin{eqnarray}
\rho(t) &=& \lambda \left\{ \sqrt{a_0 \cosh^{-2}\left[ H_0(t-t_0)
\right] - b_0 \cosh^{-4}\left[ H_0(t-t_0) \right] + c_0} - 1
\right\}, \qquad B<0, \\
p(t) &=& \frac{k_0 \cosh^{-2}\left[ H_0(t-t_0) \right] - u_0
\cosh^{-4}\left[ H_0(t-t_0) \right]}{\sqrt{ a_0 \cosh^{-2}\left[
H_0(t-t_0) \right] - b_0 \cosh^{-4}\left[ H_0(t-t_0) \right] +
c_0}}  \nonumber \\
&-& \lambda \left\{ \sqrt{a_0 \cosh^{-2}\left[ H_0(t-t_0) \right]
- b_0 \cosh^{-4}\left[ H_0(t-t_0) \right] + c_0} - 1 \right\},
\qquad B<0,
\end{eqnarray}
\begin{eqnarray}
\rho(t) &=& \lambda \left\{ \sqrt{\frac{6K}{\lambda k_4^2} \exp
\left[ -2H_0(t-t_0) \right] - \frac{12{\cal U}_0}{\lambda^2 k_4^4}
\exp\left[ -4H_0(t-t_0) \right] + c_0} - 1 \right\}, \qquad B=0,\\
p(t) &=& \frac{(2K/k_4^2) \exp\left[ -2H_0(t-t_0) \right] -
(8{\cal U}_0/\lambda k_4^2) \exp\left[ -4H_0(t-t_0) \right]}
{\sqrt{ (6K/\lambda k_4^2) \exp\left[ -2H_0(t-t_0) \right] - (12
{\cal U}_0/\lambda^2 k_4^4) \exp\left[ -4H_0(t-t_0) \right] +
c_0}} \nonumber \\
&-& \lambda \left\{ \sqrt{(6K/\lambda k_4^2) \exp\left[
-2H_0(t-t_0) \right] - ( 12{\cal U}_0/\lambda^2 k_4^4) \exp\left[
-4H_0(t-t_0) \right] + c_0} - 1 \right\}, \qquad B=0,
\label{solutf}
\end{eqnarray}
where we denoted $a_0=6 K H_0^2/\lambda k_4^2 |B|, \, b_0=12{\cal
U}_0 H_0^4/\lambda^2 k_4^4 |B|^2, \, c_0=1- 2\Lambda/\lambda k_4^2
+ 6H_0^2/\lambda k_4^2, \, k_0=2KH_0^2/k_4^2 |B|$ and $u_0=8{\cal
U}_0 H_0^4/\lambda k_4^4 |B|^2$.

An important observational quantity is the deceleration parameter,
defined as $q=(d/dt)H^{-1}-1$, given by
%\begin{eqnarray}
%q &=& - \tanh^2 \left[ H_0(t-t_0) \right], \qquad B>0, \\
%q &=& - \coth^2 \left[ H_0(t-t_0) \right], \qquad B<0, \\
%q &=& -1, \qquad\qquad B=0.
%\end{eqnarray}
\begin{equation}
q = \left\{ \begin{array}{ll}
 - \tanh^2 \left[ H_0(t-t_0) \right], & \qquad B>0, \\
 - \coth^2 \left[ H_0(t-t_0) \right], & \qquad B<0, \\
 -1, &\qquad B=0.
 \end{array} \right.
\end{equation}

There are two distinct types of behavior for the brane-world type
cosmological models described by Eqs.
(\ref{solut1})-(\ref{solutf}). For the first model, with the scale
factor having a singularity at the origin, with an appropriate
choice of the parameters, the energy density can be scaled to zero
at the beginning of the cosmological evolution, corresponding to
$t=t_0$. Then, for times $t>t_0$, the energy density is an
increasing function of time and reaches a maximum value at
$t=t_{\max}$. For time intervals $t>t_{\max}$ the energy density
is a decreasing function of time. Hence this model can be used to
describe matter creation on the brane, a phenomenon which is
entirely due to the presence of the dark matter term. If the dark
energy term ${\cal U}=0$, then, due to the singular behavior of
the scale factor, the initial energy density and pressure of the
cosmological fluid are infinite at the beginning of the
cosmological evolution, $\rho, p \to \infty$ for $a \to 0$.

For the second and third class of models, the scale factor, energy
density and pressure are all finite at $t=t_0$. Hence in this case
the brane is filled with an initial cosmological fluid. Depending
on the numerical choice of the parameters, there are also two
types of cosmological behavior, with the energy density reaching
its maximum value at the initial moment or at a time
$t=t_{max}>t_0$. In the first case for times larger than the
initial time, the energy density and the pressure are
monotonically decreasing functions for all times $t>t_0$, while in
the second case $\rho$ and $p$ are monotonically decreasing
functions of time only for $t>t_{max}$.

The behavior of the energy density and pressure is represented,
for all three cosmological models and for a particular choice of
the numerical values of the parameters, in Figs. 2 and 3.

%% REVTEX4
%%%%%%%%%%%%%%%%%%%%%%%%%%%%%%%%%%%%%%%%%%%%%%%%%%%%%%%%%%%%%%%
\begin{figure}
\includegraphics[width=10cm]{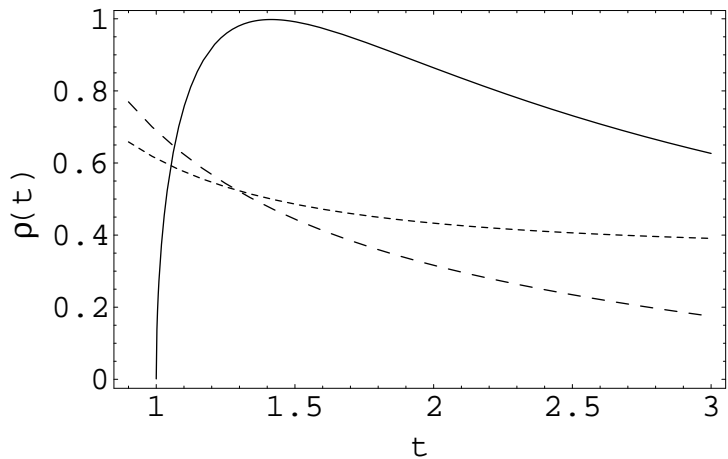}
\caption{Time evolution of the energy density $\rho(t)$ for the
brane world cosmological model, for the three classes of
solutions: first class ($a \sim \sinh\left[ H_0(t-t_0) \right]$)
(solid curve), second class ($a \sim \cosh\left[ H_0(t-t_0)
\right]$) (dotted curve) and third class ($a = \exp\left[
H_0(t-t_0) \right]$) (dashed curve). We have normalized the
parameters so that $2K=k_4^2, \, 6K/\lambda k_4^2=1, \, 12{\cal
U}_0/\lambda^2 k_4^4=1$ and $2( 3H_0^2-\Lambda )=\lambda k_4^2$.
For the sake of presentation the curves have been rescaled with
different factors.}
\end{figure}
%%%%%%%%%%%%%%%%%%%%%%%%%%%%%%%%%%%%%%%%%%%%%%%%%%%%%%%%%%%%%%%

%% REVTEX4
%%%%%%%%%%%%%%%%%%%%%%%%%%%%%%%%%%%%%%%%%%%%%%%%%%%%%%%%%%%%%%%
\begin{figure}
\includegraphics[width=10cm]{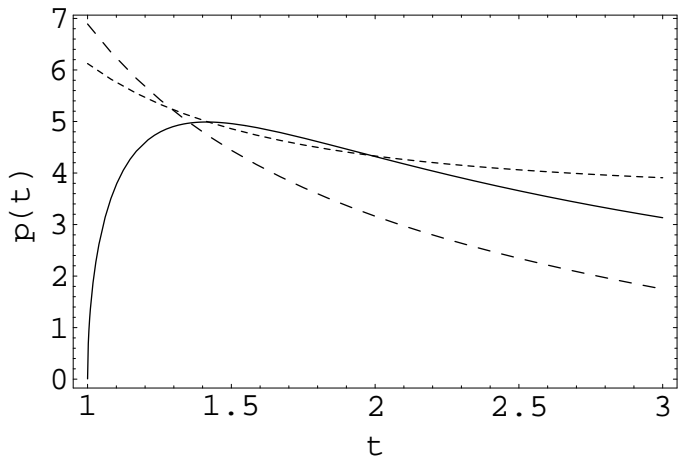}
\caption{Time evolution of the pressure $p(t)$ of the cosmological
fluid for the brane world cosmological model, for the three
classes of solutions: first class ($a \sim \sinh\left[ H_0(t-t_0)
\right]$) (solid curve), second class ($a \sim \cosh\left[
H_0(t-t_0) \right]$) (dotted curve) and third class ($a =
\exp\left[ H_0(t-t_0) \right]$) (dashed curve). We have normalized
the parameters so that $2K=k_4^2, \, 6K/\lambda k_4^2=1, \,
12{\cal U}_0/\lambda^2 k_4^4=1$ and $2( 3H_0^2-\Lambda )=\lambda
k_4^2$. For the sake of presentation the curves have been rescaled
with different factors.}
\end{figure}
%%%%%%%%%%%%%%%%%%%%%%%%%%%%%%%%%%%%%%%%%%%%%%%%%%%%%%%%%%%%%%%

The equation of state of matter is presented in Fig. 4.

%% REVTEX4
%%%%%%%%%%%%%%%%%%%%%%%%%%%%%%%%%%%%%%%%%%%%%%%%%%%%%%%%%%%%%%%
\begin{figure}
\includegraphics[width=10cm]{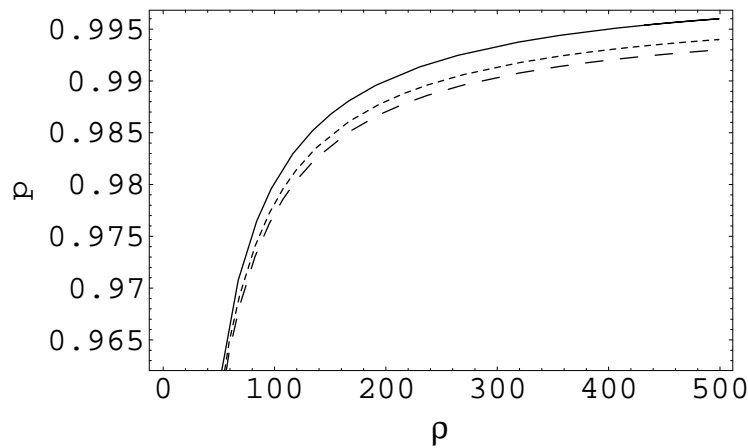}
\caption{Equation of state of the cosmological fluid for the brane
world cosmological model, for the three classes of solutions:
first class ($a \sim \sinh\left[ H_0(t-t_0) \right]$) (solid
curve), second class ($a \sim \cosh\left[ H_0(t-t_0) \right]$)
(dotted curve) and third class ($a = \exp\left[ H_0(t-t_0)
\right]$) (dashed curve). We have normalized the parameters so
that $2K=k_4^2, \, 6K/\lambda k_4^2=1, \, 12{\cal U}_0/\lambda^2
k_4^4=1$ and $2( 3H_0^2-\Lambda )=\lambda k_4^2$. For the sake of
presentation the curves have been rescaled with different
factors.}
\end{figure}
%%%%%%%%%%%%%%%%%%%%%%%%%%%%%%%%%%%%%%%%%%%%%%%%%%%%%%%%%%%%%%%

For the given choice of parameters, the pressure-energy density
dependence is generally non-linear, with a dependence which can be
approximated for small densities by a linear function, $p \sim
\rho$. In this region the effects of the contribution from the
extra-dimensions can be neglected. For high densities the
non-linear effects become important, showing that the quadratic
effects due to the effects of the five-dimensional bulk also
modify the equation of state of the matter.

An important condition that must be satisfied by any realistic
equation of state of dense matter is the requirement that the
speed of sound $c_{s}=(dp/d\rho)^{1/2}$ be smaller or equal to the
speed of light, $c_{s} \leq 1$. The time variation of the speed of
sound in the cosmological fluid with equation of state given by
Eq. (\ref{eqstate}) is represented in Fig. 5.

%% REVTEX4
%%%%%%%%%%%%%%%%%%%%%%%%%%%%%%%%%%%%%%%%%%%%%%%%%%%%%%%%%%%%%%%
\begin{figure}
\includegraphics[width=10cm]{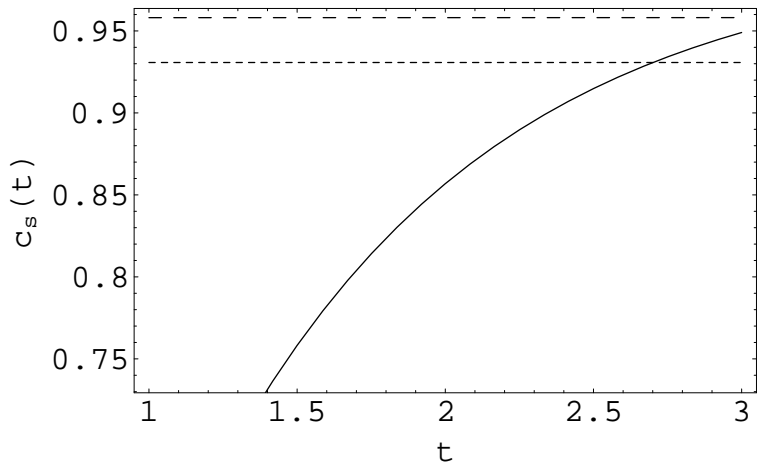}
\caption{Time variation of the speed of sound
$c_{s}=(dp/d\rho)^{1/2}$ in the cosmological fluid for the three
classes of solutions: first class ($a \sim \sinh\left[ H_0(t-t_0)
\right]$) (solid curve), second class ($a \sim \cosh\left[
H_0(t-t_0) \right]$) (dotted curve) and third class ($a =
\exp\left[ H_0(t-t_0) \right]$) (dashed curve). We have normalized
the parameters so that $2K=k_4^2, \, 6K/\lambda k_4^2=1, \,
12{\cal U}_0/\lambda^2 k_4^4=1$ and $2( 3H_0^2-\Lambda )=\lambda
k_4^2$.}
\end{figure}
%%%%%%%%%%%%%%%%%%%%%%%%%%%%%%%%%%%%%%%%%%%%%%%%%%%%%%%%%%%%%%%

For the given range of parameters and in the considered time
interval, the condition $c_{s}\leq 1$ is satisfied. During this
period the speed of sound is a rapidly increasing function of time
for the first solution and a slowly increasing function of time
(almost a constant) for the second and third class of solutions.
But for other choices of the numerical values of the constants or
different time intervals this condition could be violated.

Since $q<0$ for all times, the evolution of the models is purely
inflationary. However, due to the corrections from the
extra-dimensions, the initial period of the inflationary phase can
be associated to an increase in the energy-density of the
Universe.

In all these cases, due to the rapid expansion of the brane-world,
there is a fast time decay of the angular velocity $\omega \sim
a^{-3}$. Since in the large time limit the scale factor is of the
form $\exp(H_0 t)$, in this brane world model one obtains an
exponential decay of the cosmic vorticity.

%%%%%%%%%%%%%%%%%%%%%%%%%%%%%%%%%%%%%%%%%%%%%%%%%%%%%%%%%%%%%%%%%%%%%%
\section{Discussions and final remarks}
%%%%%%%%%%%%%%%%%%%%%%%%%%%%%%%%%%%%%%%%%%%%%%%%%%%%%%%%%%%%%%%%%%%%%%
In the present paper we have considered the evolution of the
rotational perturbations in the framework of brane world
cosmology. As a first step we have obtained the basic dynamical
equation governing the temporal variation and spatial distribution
of the metric perturbation function $\Omega(t,r)$. This equation
also contains the angular velocity of the slowly rotating matter.
From the consistency condition of the rotation equation one can
obtain some restrictions on the admissible mathematical form of
$\omega$. As a general result we have shown that in the presence
of dark energy, describing the effects of the bulk on the brane,
rotational perturbations always decay for the slowly rotating
brane. But for a vanishing dark energy term, rotational
perturbations do not decay for very high density stiff matter with
$\gamma=2$, except for the case of the perfect dragging. Hence the
dark energy term could have play an important role in suppressing
the vorticity of the very early Universe.

In order to find the time evolution of the four-velocities of the
particles on the brane we consider the dynamics of a test particle
in the perturbed metric (\ref{R6}). The equations of motion are
$du^\mu/ds+\Gamma^\mu{}_{\nu\lambda} u^\nu u^\lambda=0$, where
$u^\mu$ are the components of the four-velocity and the
Christoffel symbols $\Gamma^\mu{}_{\nu\lambda}$ are computed from
the metric. To simplify the calculations we consider only the
first order corrections to the metric in $\Omega$ and assume that
test particles have small velocities, thus retaining only terms
which are linear in velocity. Consequently we obtain
\begin{equation}
u^0 = 1, \quad \frac{du^1}{dt} = - \frac{2\dot{a}}{a} u^1, \quad
\frac{du^2}{dt} = 0, \quad \frac{ du^3}{dt} = \frac{2\dot{a}}{a}
\Omega + \dot{\Omega} - \frac{2\dot{a}}{a} u^3. \label{vel}
\end{equation}

By integrating Eqs. (\ref{vel}) we find
\begin{equation}
u^1 = \frac{u_0^1}{a^2(t)}, \quad u^2 = u_0^2, \quad
u^3=\frac{u_0^3}{a^2(t)} + \frac{A(r)}{a^3(t)},
\end{equation}
with $u_0^i,\, i=1,2,3$ constants of integration. $u_0^2$ can be
set to zero without altering the physical structure. In the limit
of large $t$, $a(t)\to\infty$ and we have $u^1=0$ and $u^3=0$. At
the time $t=t_{\infty}$ the test particle will have shifted in
azimuth by $\Delta\varphi=\int_0^{t_{\infty}} u^3 dt$, having an
angular velocity $\omega=\frac{\Delta\varphi}{\Delta t}$.
Present-day observations impose a strong restriction on the
numerical value of the angular velocity of the Universe of the
form $\omega(t_{\infty})<10^{-15} \text{years}^{-1}$. Hence in
realistic cosmological models the angular velocity must tend to
zero in the large time limit.

The condition of the mathematical consistency of the rotation
equation also leads to a specific brane world cosmological model,
with the equation of state of the matter given in a parametric
form. In the high density regime the pressure-energy density
dependence is nonlinear. The presence of the dark energy term has
major implications on the time evolution of physical parameters.
Due to the presence of ${\cal U}$, the energy density and pressure
of the matter on the brane have a maximum. The increase in the
energy density from zero to a maximum value can model the energy
transfer from the bulk to the zero-density brane, thus leading to
the possibility of a phenomenological description of matter
creation on the brane. If the dark energy term is zero, the
singularity in the scale factor is associated to a singular
behavior of $\rho$ and $p$. In the large time limit, $a\to\infty$
and $\rho\to\lambda \left( \sqrt{1 - 2\Lambda/\lambda k_4^2 +
6H_0^2/\lambda k_4^2} - 1 \right)$, with the pressure $p \to -\rho
$. Since for ordinary matter $p$ cannot be smaller than zero, it
follows that in this limit $\rho$ must be zero, condition which
leads to $H_0=\sqrt{\Lambda/3}$ and $a=\exp(H_0 t)$. Therefore the
de Sitter solution is an attractor for this brane world model.
Independently on the initial state, the Universe ends in an
inflationary phase.

On the other hand one must point out that the curvature of the
brane, described by the parameter $k=-1,0,+1$ does not play any
significant role in the time evolution of this brane-world model.
Since $k$ is absorbed in the arbitrary constant $B$ (which in fact
is the most important parameter of the model), the dynamics of the
matter on the brane is independent on the three-dimensional
geometry of the Universe. However, the knowledge of the spatial
distribution of the angular velocity could, at least in principle,
allow for the determination of the separation constant $K$. If the
numerical value and sign of $K$ would be known, then the effects
of the curvature of the brane on the temporal dynamics could also
be obtained. But due to the exponential increase in the scale
factor, the Universe ends in a non-rotating state with a flat
geometry.

Since the equation of state of matter is unusual (even in the
extreme limit of high densities the equation of state of ordinary
matter is still linear with $\rho =p$), it is natural to assume
that in this cosmological model the initial matter content of the
Universe on the brane consists from a field rather than ordinary
matter. The best candidate is a scalar field $\phi$, with
$\rho=\rho_{\phi}=\frac{\dot{\phi}^{2}}{2}+V(\phi) \geq 0$ and
$p=p_{\phi}=\frac{\dot{\phi}^{2}}{2}-V(\phi)$, where $V(\phi) \geq
0$ is the self-interaction potential. Then the time evolution of
the scalar field can be immediately obtained from
$\dot{\phi}(t)=\sqrt{2(\rho+p)}$, while the potential is given by
$V(t)=(\rho-p)/2$. With the use of Eqs.(\ref{solut1})-(\ref
{solutf}) one can easily obtain the time dependence of the scalar
field and scalar field potential for each class of solutions. The
scalar field drives the Universe into an inflationary era. In this
case solutions with negative pressure are also allowed. In the
limit of large times, since the field satisfies the equation of
state $\rho+p=0$, we obtain $\dot{\phi}=0$ and
$\phi=\text{constant}$. For the scalar field potential we find
$V\to \rho $, and, in the de Sitter limit, it follows $V=0$. Hence
$V$ does not give any contribution to the cosmological constant
$\Lambda$.

%%%%%%%%%%%%%%%%%%%%%%%%%%%%%%%%%%%%%%%%%%%%%%%%%%%%%%%%%%%%%%%%%%%%%%
\section*{Acknowledgments}
%%%%%%%%%%%%%%%%%%%%%%%%%%%%%%%%%%%%%%%%%%%%%%%%%%%%%%%%%%%%%%%%%%%%%%
This work is supported in part by the National Science Council
under the grant numbers NSC90-2112-M009-021 and
NSC90-2112-M002-055. The work of CMC is also supported by the
Taiwan CosPA project and, in part, by the Center of Theoretical
Physics at NTU and National Center for Theoretical Science. CMC is
grateful to the hospitality of the KIAS (Korea) where part of this
work is done.

\end{document}